\documentclass[12pt,nohyper,notoc]{JHEP}

\usepackage{amsmath,amssymb,cite} 
\input epsf
	
\font\zfont = cmss10 
\newcommand\ZZ{\hbox{\zfont Z\kern-.4emZ}}
\def\inbar{\vrule height1.5ex width.4pt depth0pt}
\def\IC{\relax\hbox{\kern.25em$\inbar\kern-.3em{\rm C}$}}

\newcommand{\EQ}[1]{\begin{equation} #1 \end{equation}}
\newcommand{\AL}[1]{\begin{subequations}\begin{align} #1 \end{align}\end{subequations}}
\newcommand{\SP}[1]{\begin{equation}\begin{split} #1 \end{split}\end{equation}}

\title{Universal Aspects of Gravity Localized on Thick Branes}

\author{Csaba Cs\'aki$^{a,}$\footnote{J. Robert Oppenheimer Fellow}, 
Joshua Erlich$^a$, Timothy J. Hollowood$^{a,b}$ and Yuri Shirman$^c$\\
$^a$Theory Division T-8, Los Alamos National Laboratory, Los Alamos,
NM 87545, USA\\
$^b$Department of Physics, University of Wales Swansea,
Swansea, SA2 8PP, UK\\
$^c$Department of Physics, Princeton University, Princeton, NJ 08544, USA\\

Email: {\tt csaki@lanl.gov, erlich@lanl.gov, pyth@skye.lanl.gov,
yuri@feynman.princeton.edu}}

\abstract{We study gravity in backgrounds that are smooth generalizations 
of the Randall\-Sundrum model, with and without scalar fields.  These 
generalizations
include three-branes in higher dimensional spaces which are not 
necessarily Anti-de Sitter far from the branes,
intersecting brane configurations and configurations involving negative
tension branes.  We show that under certain
mild assumptions there is a universal equation for the gravitational
fluctuations.  We study both the graviton ground state and the continuum of
Kaluza-Klein
modes and we find that the four-dimensional gravitational mode is localized
precisely when the effects of the continuum modes decouple at 
distances larger than the fundamental Planck scale. The decoupling is
contingent only on the long-range behaviour of the metric from the
brane and we find a universal form for the corrections to Newton's
Law. We also comment on the possible contribution of resonant modes.
Given this, we find
general classes of metrics which maintain localized four-dimensional gravity.
We find that three-brane metrics in five dimensions can 
arise from a
single scalar field source, and we rederive the BPS type
conditions without any {\em a priori\/} assumptions regarding the form of the
scalar potential. We also show that a single scalar field
cannot produce conformally-flat locally intersecting brane configurations or a
$p$-brane in greater than $(p+2)$-dimensions.}

\preprint{{\tt hep-th/0001033}\\PUPT-1912}

\begin{document}

\section{Introduction}
\label{sec:intro}

The proposal of Randall and Sundrum (RS) \cite{RS,RS2} 
to localize gravity in the vicinity of a brane 
with non-vanishing tension in anti-de Sitter (AdS) space has recently 
attracted enormous attention (see, for example 
\cite{noncompact,Gogberashvili}, for previous relevant work,
\cite{ADDK,CS,Ann,SeanMark,Nam,Kaloper,LR,Wise,GW,CGS2,Kraus,Kaloper2,others,
CK,Erich,Nima2,Nelson2,Kogan,Gabadadze,Chaichian}, 
for more recent generalizations, 
\cite{oldCvetic,cveticdw,DFGK,Chamblin,cvetic,Gremm}, for work on 
smooth brane scenarios,
\cite{Kehagias,Verlinde,skenderis,Kostas,Gubser,Youm,Kallosh:2000tj}, for embeddings in
string theory and supergravity,
\cite{CHR,EHM,Japanese,Aref'eva:1999,CGR,Ivanov:1999,Gauss,Chamblin2,Dick,Leontaris:1999iw},
for the general relativity aspects and finally
\cite{csaba2,Tye2,BDEL,Keith,DHR}, for cosmological and
phenomenological aspects).
RS found that in a setup with a single brane,
a negative bulk cosmological constant and a single large extra dimension
(with the cosmological constant and brane tension tuned such that the 
effective four-dimensional cosmological constant vanishes) the solution to Einstein's 
equation results in a single graviton zero mode, which is a consequence
of the unbroken four-dimensional Poincar\'e invariance, and a continuum
of Kaluza-Klein (KK) modes. Normally
the presence of these continuum modes would render a setup like this
unrealistic due to the large deviation from Newton's Law the low energy 
continuum modes tend to induce. 
However, RS found that due to the suppression of the 
wavefunctions of the continuum modes close to the brane, their contribution
to the Newton potential is highly suppressed, and therefore a realistic model
with uncompactified extra dimensions could be built. This model has 
been generalized in \cite{ADDK} to models with intersecting branes 
with more than one uncompactified extra dimension, and also to include
brane junctions \cite{CS,Ann}.

The branes in the RS setup and its generalizations mentioned above are included
as static point-like external sources in the extra dimensions, with no 
dynamics to produce them. As was done in \cite{DFGK,Chamblin,cvetic,skenderis,cveticdw},
one can find solutions to Einstein's equation coupled to a 
single scalar field, where the scalar creates a domain wall---a
``thick brane''---while the metric away from the brane asymptotes
to a slice of AdS$_5$. Such domain wall solutions are obtained 
if the scalar potential originates from a superpotential (although as
recently discussed in \cite{Kallosh:2000tj} this does not necessarily imply
that the theory is embeddable into a five-dimensional supergravity theory). In this
case the solutions found in \cite{DFGK,Chamblin,cveticdw} originate from a BPS
equation. These domain walls were first found in 
\cite{oldCvetic}. It has been shown in \cite{DFGK,Chamblin,cveticdw} that, just like
for the case of the infinitely thin branes of RS, there is a single
normalizable graviton bound-state with zero energy. A particularly
nice example of this sort has been recently worked out in detail
in \cite{Gremm}. Similar BPS equations for intersecting domain walls in 
more than one extra dimension were found in \cite{SeanMark,Nam};
however, no explicit solutions to these equations are known yet.

In this paper we study generic properties of localized gravity on thick
branes. In the first part of the paper we consider thick branes in one
extra
dimension and then generalize to an arbitrary number of extra dimensions.
Instead of starting with a coupled gravity-scalar system, as in
\cite{DFGK,Chamblin}, we ``smear''
the RS solution and its generalizations in such a way that the non-dynamical
source terms correspond to 
a smeared (thick) brane in the background of a slowly varying 
negative bulk cosmological constant.
We examine the spectrum of graviton modes and find
necessary and sufficient conditions for such backgrounds to localize
gravity on the branes.
Besides general arguments about the ground state (some of which appear in 
\cite{DFGK,Gremm}), we also examine the
behavior of the continuum modes. For a generic study, we use the 
WKB approximation for the ``volcano-type potential'', which hints that
when the metric falls off slowly enough from the brane
the soft KK modes are suppressed sufficiently inside the brane so that
their corrections to the Newton's Law are negligible.  In a more
restricted set of generalizations of the RS solution we calculate this
suppression more rigorously, and find qualitative agreement with the WKB
result.  This leads to necessary conditions of the quantum mechanical 
potential, and hence the background metric, in order 
for the KK modes to decouple. We find that the potential 
at large distances must fall off no faster than in the asymptotically
AdS case in order for the KK modes to make a small contribution to 
Newton's Law.  This requirement is equivalent to demanding normalizability of
the ground state graviton wavefunction. 
We also comment on the possible contributions of
``quasi-bound-states''---resonant modes in the continuum spectrum 
whose wavefunctions are not suppressed at the location of the brane---and
study their significance in a toy model.  We next
show how to generalize our results to situations in more than five
dimensions. These scenarios could describe, for example,
three-branes in more than
five dimensions or higher-dimensional intersecting branes with a
four-dimensional intersection. In the latter case,
the thick brane
background could be given by an appropriate smearing of the intersecting brane 
solution of \cite{ADDK}, and again we find conditions on the 
long-distance behavior of the background
metric in order for there to be localized gravity on the brane intersection.

We also study the relevance of background fields
that create the branes.  We find that the stress-tensor source terms
for a general smearing of the RS solution can be 
obtained from a single scalar field, and we rederive the same BPS-type
equations for this scalar field as \cite{DFGK,Chamblin,cveticdw}. This
provides a particularly simple derivation of the BPS equations without
an {\it a priori\/} assumption about the form of the scalar potential, and 
also emphasizes that this is the most general solution with a single scalar 
field.
In the case of branes in higher dimensions the situation is more complicated.
Contrary to the case of one extra
dimension, we find that it is impossible to generate the desired background
metric or sources of the stress tensor with a single scalar field.
Nevertheless, the properties of the graviton in such backgrounds are 
studied in the same way as for the case of one extra dimension.

\section{Backgrounds with four-dimensional Poincar\'e invariance}

In general, we are interested in $d$-dimensional backgrounds which
have a four-dimensional Poincar\'e symmetry (the restriction to four dimensions
is unnecessary, but is the case most relevant for phenomenology):
\EQ{
ds^2\equiv g_{\mu\nu}dx^\mu\,dx^\nu=e^{-A(z)}\eta_{ab}\,dx^a\,dx^b+g_{ij}(z)dz^i\,dz^j\ .
\label{metric}
}
Here, $x^\mu=(x^a,z^i)$, where $x^a$, for $a=0,\ldots,3$, 
are the usual coordinates of four-dimensional Minkowski space
and $z^i=x^{i+3}$, for $i=1,\ldots,n$, are the coordinates on the $n=(d-4)$-dimensional
transverse space.\footnote{In our conventions the metric $g_{\mu\nu}$
has signature $(+,-,-,-,\ldots)$.}
We will assume that $A(0)=0$, so that the
four-dimensional metric at the origin in the transverse space is
canonically normalized. In this present work we
will concentrate, for the most part, on a
more restricted set of metrics which are conformally flat; that is of
the form
\begin{equation}
ds^2=e^{-A(z)}\big(\eta_{ab}\,dx^a\,dx^b-dz^i\,dz^i\big)\ ,
\label{confflat}\end{equation}
with a suitable choice of coordinates.
Notice that when $d=5$, an arbitrary metric of the form \eqref{metric}
is conformally flat, so in that case \eqref{confflat} is perfectly general. 

We will be interested, amongst other things, in smooth
versions of the RS metric in five dimensions  discussed in
\cite{RS}. The metric is usually written in the form
\EQ{
ds^2=e^{-A(r)}\eta_{ab}\,dx^a\,dx^b-dr^2
\label{rsmetric}
}
with $A(r)=2k|r|$,
but can be written in conformally-flat form 
with $A(z)\equiv A(r(z))=2\log(k|z|+1)$.
One simple way to introduce thick brane is by smoothing out
RS ansatz. For example, one can make a substitution $|r| \rightarrow
\mu^{-1}\log\cosh(\mu r)$, although for practical purposes it is more
convenient to smooth in the $z$-basis:
\EQ{
|z| \rightarrow \frac{1}{\mu} \log \cosh (\mu z) \, .
}
Here $\mu$ is an independent parameter which determines the
thickness of the brane. In the RS limit, $\mu \gg k$, we
expect that any additional fields will be localized near
$z=0$, which will be assumed in much of the following
discussion (we will comment on subtleties associated with
smearing of the matter fields over the thickness of the brane
later). 
On the other hand, for simplicity it is convenient to study the
behaviour of the gravitational modes in a different limit
$\mu \sim k$, where $A(z)$ depends on only one scale $k$,
and we will do so in most of our examples.
In this case the equivalent smoothing in conformally-flat
coordinates can be written as $A(r)=2\log\cosh(kr)$, or equivalently
$A(z)=\log(k^2z^2+1)$, \cite{Gremm}.
This metric
approaches the AdS form asymptotically for $|z|\gg1/k$.
However, we will also consider metrics which
are more general and not necessarily asymptotic to an AdS space.

We first point out that smoothing of the RS solution can be performed without
the addition of matter fields.  
Consider the five-dimensional case where the domain wall is generated by an 
explicit
position dependent term in the gravity action, much in the spirit of the
original Randall-Sundrum scenario.  (We will later study gravity in the 
background
of branes created by scalar fields, and find that the supersymmetric
potential introduced in \cite{DFGK,Chamblin} appears naturally in the solution
of the field equations.)  In the absence of fields other
than gravity, we study the action
\begin{equation}
S=-\int d^5x \,\left[\sqrt{g}\left(\kappa^{-2}R+\Lambda(r) \right)+\sqrt{g_{(4)}}\,V(r)
\right]\ . 
\label{5daction}
\end{equation}
Here $g=|{\rm det}\,g_{\mu\nu}|$, $g_{(4)}=|{\rm det}\,g_{ab}|$ is the 
determinant of the induced metric on the domain wall and $
\kappa^2=M_*^{-3}$, where $M_*$ is the fundamental Plank scale in five
dimensions.  
The function $V(r)$ will approximate a delta function 
which generates the domain wall, and
$\Lambda(r)$ is roughly 
constant away from the domain wall and corresponds to the bulk cosmological 
constant.  This action gives rise to a stress tensor 
\begin{equation}
T_{\mu\nu}=\tfrac{1}{2}\left(\Lambda(r)\,g_{\mu\nu}+V(r)\,g_{ab}\,
\delta_\mu^a\, \delta_\nu^b\right)\ .
\label{stform}
\end{equation}
We note that any stress tensor which is four dimensional Lorentz
covariant can be decomposed in this way, and therefore is derivable from
an action of the form \eqref{5daction}.
We are interested in actions which admit solutions of the form 
\eqref{rsmetric}.
The ``bulk cosmological constant'' and ``brane tension'' are then 
determined to be 
\EQ{
\Lambda(r)=-3\kappa^{-2}A^\prime(r)^2\ ,\qquad
V(r)=3\kappa^{-2}A^{\prime\prime}(r)\ . \label{lambdav}
}
For $A(r)=2k|r|$, this reproduces the original Randall-Sundrum system 
\cite{RS}. By choice of a static background metric \eqref{rsmetric} the
four-dimensional effective cosmological constant must vanish, and the
fine tuning of cosmological constant and brane tension is replaced by
\eqref{lambdav}.  Indeed, one can integrate out the extra dimension and
check that the action \eqref{5daction} 
vanishes for the background metric \eqref{rsmetric}, which is equivalent
to vanishing of the four-dimensional cosmological constant.

We should comment that because this analysis does not depend on what type
of field creates the domain walls, we can study solutions in which the
domain walls have
negative tension.  For example, we can study smooth versions of the RS solution
to the hierarchy problem \cite{RS2}, in which the transverse 
direction is compactified on an orbifold $S^1/{\mathbb Z}_2$.  Branes are stuck
at the two fixed planes of the orbifold action, one of which has positive
tension and the other negative tension.  In the RS scenario the metric has
the form \eqref{rsmetric} with $A(r)=2k|r|$, $-r_c<r<r_c$, where 
it is understood that $r$ is periodic over an interval $2r_c$.  In order
to make the periodicity explicit, we can study a multiple cover of the
circle, with  
\begin{equation}
A(r)=\sum_{n\in{\mathbb Z}} k\left[r_c+(-1)^n\left(2r-(2n+1)\,
r_c\right)\right]\,\theta(r-nr_c)\,\theta((n+1)r_c-r)\ ,
\label{A} \end{equation}
where $\theta(r)$ is the usual step function.
The brane tension is related to $A^{\prime\prime}(r)$, which has delta
function singularities with positive coefficient for $r=0$ mod $2r_c$
and negative coefficient for $r=r_c$ mod $2r_c$.
As before the solution can be smoothed by an appropriate smearing of the
$\theta$-functions; for example, 
\begin{equation}
\theta(x)\rightarrow \tilde\theta(x)\equiv \tfrac{1}{2}(\tanh(\mu x)+1)\ ,
\end{equation}
where $\mu$ is, as described previously, a parameter 
which characterizes the thickness of the brane. 
An alternative smoothing procedure which may be useful for numerical
calculation is a truncation of the Fourier expansion of the sawtooth:
\EQ{
A(r)\approx \frac{r_c}{2}
-\frac{4r_c}{\pi^2}\sum_{0\leq n<N}\frac{1}{(2n+1)^2}
\cos(\pi(2n+1)r/r_c)\ .
}

We will also consider backgrounds that can
be interpreted as three-branes embedded in spaces of dimension
$d>5$ and also backgrounds which can be
interpreted as intersections of higher dimensional branes with a
four-dimensional intersection. In the latter cases, 
the question for these kinds of
backgrounds is whether four-dimensional gravity is 
localized on the intersection.
For example, we will consider smooth versions of the multi-dimensional
patched AdS space with metric \cite{ADDK}
\EQ{
ds^2=\frac1{(1+k\sum_{i=1}^{n}|z^i|)^2}
\big(\eta_{ab}\,dx^a\,dx^b-dz^idz^i\big)\ .
\label{mdrsmetric}
}
This metric represents $n=d-4$ intersecting $(d-1)$-branes,
which mutually intersect in four dimensions.

\section{Universal Aspects}\label{sec:univ}

\subsection{Gravitational Fluctuations}\label{sec:fluct}

In this section we derive a universal equation for the 
effective four-dimensional gravitational 
fluctuations in the conformally-flat backgrounds described in the last
section. The more demanding case of the general background \eqref{metric} will be
discussed separately in Sec.~\ref{sec:gen}.
This means that we study fluctuations of the metric \eqref{confflat}
of the form
\begin{equation}
ds^2=e^{-A(z)}\big((\eta_{ab}+h_{ab}(x,z))\,dx^a\,dx^b-
dz^i\,dz^i\big)\ .
\label{ddpert}\end{equation}
It will be convenient to define $h_{\mu\nu}$ to be a fluctuation whose only non-zero
components are $h_{ab}$. We will use the transverse traceless gauge
for these fluctuations, {\em i.e.}~$\partial_\mu\,h^{\mu\nu}=0$ and 
$h^\mu_\mu=0$.  We should note that there may be additional fluctuations
not of the form \eqref{ddpert} in transverse traceless gauge, but we will not
comment on such modes here.  

Since the metric \eqref{confflat}
is manifestly conformally flat, it is convenient to use the the
general form of the Einstein tensor for metrics of the form $g_{\mu\nu}=e^{-A}\tilde{g}
_{\mu\nu}$ (see for example \cite{Wald}): 
\EQ{
G_{\mu\nu}=\widetilde{G}_{\mu\nu}+\tfrac{d-2}{2}\Big[\tfrac12
\widetilde{\nabla}_\mu A\,
\widetilde{\nabla}_\nu A+\widetilde{\nabla}_\mu\widetilde{\nabla}_\nu
A-\tilde{g}_{\mu\nu}\big(\widetilde{\nabla}_\rho\widetilde{\nabla}^\rho
A
-\tfrac{d-3}{4}\,\widetilde{\nabla}_\rho A\,\widetilde{\nabla}^\rho
 A\big)\Big]\ ,
\label{conformal}
}
where indices are raised and lowered with $\tilde{g}_{\mu\nu}$ in this 
context. Using the form of the Einstein tensor for linear perturbations about flat
spacetime \cite{Wald}, 
\begin{equation}
\delta\widetilde{G}_{\mu\nu}=\partial^\rho\partial_{\left(\nu\right.}
h_{\left.\mu\right)\rho}-\tfrac{1}{2}
\partial^\rho\partial_\rho
\,h_{\mu\nu}-\tfrac{1}{2}\partial_\mu\partial_\nu 
h^\rho
_\rho-\tfrac{1}{2}\eta_{\mu\nu}\left(\partial^\rho\partial^\kappa\,h_{\rho
\kappa}
-\partial^\rho\partial_\rho\, h^\kappa_\kappa\right),
\end{equation}
the linearized perturbation of the Einstein tensor \eqref{conformal}
becomes, 
\SP{
\delta G_{\mu\nu}&=\partial^\rho\partial_{\left(\nu\right.}
h_{\left.\mu\right)\rho}-\tfrac{1}{2}\underline{\partial^\rho
\partial_\rho \,h_{\mu\nu}}
-\tfrac{1}{2}\partial_\mu\partial_\nu\,  h^\rho
_\rho-\tfrac{1}{2}\eta_{\mu\nu}\left(\partial^\rho\partial^\kappa\,h_{\rho
\kappa}
-\partial^\rho\partial_\rho\, h^\kappa_\kappa\right)\\ 
&-\tfrac{d-2}{2}\big[\tfrac{1}{2}\eta^{\rho\kappa}(\partial_{\mu}\,h_{\nu\rho}
+\partial_\nu\,h_{\mu\rho}-\underline{\partial_\rho\,h_{\mu\nu}})\,
\partial_\kappa\,A\\
&+\underline{h_{\mu\nu}\partial_\rho\partial^\rho A}
+\eta_{\mu\nu}\left(
-\partial_\rho\,h^{\rho\kappa}\,\partial_\kappa A
+\tfrac{1}{2}\partial_\rho\,h^\kappa_\kappa\,\,\partial^\rho A
-h^{\rho\kappa}\,\partial_\rho\partial_\kappa A \right)\\
&-\tfrac{d-3}{4}\big(\underline{h_{\mu\nu}\,\partial_\rho
 A\,\partial^\rho
 A}-\eta_{\mu\nu}\,h^{\rho\kappa}\partial_\rho A\,
\partial_\kappa A \big)\big]\ .
}
Only the terms which are underlined are actually non-zero. The other
terms either vanish due to the gauge conditions or they vanish because
$h_{\mu\nu}$ only has non-zero components $h_{ab}$ and, moreover, $A$ is a
function of the $\{z^i\}$ only.

The next question concerns the variation of the stress tensor. 
For any action of the form \eqref{5daction}
\EQ{
\delta T_{\mu\nu}=T_\mu^\kappa\,h_{\kappa\nu}\ ,
\label{tst}
}
which is automatically symmetric. Later we will show that this
transformation property remains valid when the background is
generated by scalar fields. From \eqref{tst}, and the
unperturbed Einstein equation, we derive
\EQ{
\delta T_{\mu\nu}=\tfrac{d-2}{2}\kappa^{-2}(\tfrac12\partial_\mu A\,
\partial^\kappa A+\partial_\mu\partial^\kappa A
)\,h_{\kappa\nu} 
+\tfrac{d-2}{2}\kappa^{-2}(-\underline{\partial_\rho
\partial^\rho A}
+\tfrac{d-3}{4}\,\underline{\partial_\rho A\,\partial^\rho A})h_{\mu\nu}\ .
}
Again only the underlined terms survive due to either the gauge
conditions or the particular choice of background. 
Both the surviving terms of $\delta T_{\mu\nu}$ 
cancel with two terms of $\delta G_{\mu\nu}$ in the graviton equation of motion
$\delta G_{\mu\nu}-\kappa^2\delta T_{\mu\nu}=0$, leaving 
\EQ{
-\tfrac{1}{2}\partial^\rho\partial_\rho\,h_{\mu\nu}
+\tfrac{d-2}{4}\,\partial^\rho A\,\,\partial_\rho\,h_{\mu\nu} =0\ .
\label{dddg}
}
If we redefine the metric perturbation so that its kinetic term has the
canonical normalization, {\em i.e.}~${h}_{\mu\nu}=e^{(d-2)A/4}\,\tilde h_{\mu\nu}$,
then the term linear in derivatives is removed:
\EQ{
-\tfrac{1}{2}\partial^\rho\partial_\rho\,\tilde{h}_{\mu\nu}+
\Big[\tfrac{(d-2)^2}{32}
\partial^\rho A\partial_\rho
A-\tfrac{d-2}{8}\partial^\rho\partial_\rho A\Big]\tilde{h}_{\mu\nu}
=0\ . 
\label{ddgraviton} 
}
We should emphasize that indices are raised and lowered here using the
flat metric $\eta_{\mu\nu}$. In addition, only the $h_{ab}$ components
of the fluctuation $h_{\mu\nu}$ are non-vanishing.

Now we use the fact that
$\partial^\rho\partial_\rho=-\square_x-\nabla_z^2$, where
$\square_x=-\eta^{ab}\partial_a\partial_b$ and $\nabla_z^2=\partial_i^2$,
and look for solutions of the form $\tilde{h}_{ab}(x,z)=\check
h_{ab}(x)\psi(z)$ with $\square_x\check
h_{ab}(x)=m^2\check h_{ab}(x)$, where $m$ is the four-dimensional Kaluza-Klein  
mass of the fluctuation. Then since $A=A(z)$ only, we have
\EQ{
-\nabla_z^2\psi(z)+
\Big[\tfrac{(d-2)^2}{16}
\nabla_zA\cdot\nabla_z
A-\tfrac{d-2}{4}\nabla_z^2A\Big]\psi(z)=m^2\psi(z)\ .
\label{mdschreq} 
}
This has the form of a Schr\"odinger equation for
the ``wavefunction'' $\psi(z)$, ``energy'' $m^2$ and potential
\EQ{
V(z)=\tfrac{(d-2)^2}{16}
\nabla_zA(z)\cdot\nabla_z
A(z)-\tfrac{d-2}{4}\nabla_z^2A(z)\ .
\label{schpot}
}

The 
fact that when expressed in terms of the variable $\tilde h_{\mu\nu}$, 
the graviton equations-of-motion \eqref{ddgraviton} have no single
derivative terms is equivalent to the fact that the action for these
fluctuations has the form of a canonical kinetic term:
\EQ{
S \thicksim \int d^dx\,\partial_\rho\tilde
h_{\mu\nu}\partial^\rho\tilde h^{\mu\nu}\ ,
\label{s}
}
where indices are contracted with the flat metric $\eta_{\mu\nu}$.  This can
be seen 
by expanding the scalar curvature about the background \eqref{confflat} and
keeping track of powers of the conformal factor in the metric.  (If we had
chosen coordinates in which the metric is not explicitly in the
conformally-flat form, then the kinetic terms
in the $x$ and $z$ directions would have had different conformal factors
multiplying them.)  The redefinition of the metric $h_{\mu\nu}=e^{(d-2)A(z)/4}
\tilde{h}_{\mu\nu}$ then absorbs the conformal factor multiplying the
kinetic terms and puts the action in the canonical form \eqref{s}.
In particular, for solutions in the form $\check h_{ab}(x)\psi(z)$,
\eqref{s} includes the term
\EQ{
\int d^nz\,\psi(z)^2\,\cdot\,\int d^4x\,\partial_c\,\check h_{ab}(x)\,
\partial^c\,\check h^{ab}(x)\ ,
}
from which we deduce that the appropriate inner-product for the
``wavefunctions'' $\psi(z)$ is the conventional
quantum mechanical one. 
(Notice that this differs from the inner-product employed in \cite{DFGK}.)

In order to calculate the strength of the four-dimensional gravitational 
coupling, it is convenient to decompose the action into a four 
dimensional part and higher dimensional parts, before rescaling the graviton.
As in \cite{RS,ADDK}, including the fundamental ($d$-dimensional) Planck scale $M_*$, which
is related to the coupling $\kappa$ via
\EQ{
\kappa^2=M^{2-d}_*\ ,
}
the action takes the form,
\begin{equation}
S\thicksim M^{d-2}_*\int d^nz\,e^{-(d-2)A/2}\, \cdot\,\int d^4x\,\sqrt{
\hat{g}_{(4)}}
\,R^{(4)} + \cdots ,
\end{equation}
where $\hat{g}_{(4)}$ is the determinant of the four-dimensional metric 
for matter perturbations about flat spacetime, and $R^{(4)}$ is the 
four-dimensional curvature scalar created by those matter perturbations.
This allows us to identify the four-dimensional Planck scale $M_4$ via,
\begin{equation}
M_4^2=M^{d-2}_*\,\int d^nz\,e^{-(d-2)A(z)/2}\ ,
\label{4dplanck}\end{equation}
and determines the four dimensional gravitational coupling $G_N\sim
M_4^{-2}$.\footnote{Note 
that the above relation relies on our choice $A(0)=0$.}

\subsection{The four-dimensional graviton}\label{sec:fdgrav}

The question of whether there is localized (four-dimensional) gravity supported in the
vicinity of the brane now becomes contingent on properties of the
quantum mechanical system described by the Schr\"odinger equation
\eqref{mdschreq}. In particular, in order to have an effective
four-dimensional theory of gravity we require that \eqref{mdschreq} admits
a normalizable zero-energy ground state.
To find this zero-energy state, we notice,
generalizing the observation of \cite{DFGK,Kostas} to higher dimensions, 
that the Schr\"odinger equation \eqref{mdschreq} can be rewritten as a 
supersymmetric quantum mechanics problem of the form,
\EQ{
Q^\dagger\cdot Q\,\psi(z)=m^2\psi(z)\ ,
\label{susy}
}
where the $n$-vector of supersymmetry charge is
\EQ{
Q=\nabla_z+\tfrac{d-2}4\nabla_zA\
,\qquad Q^\dagger=-\nabla_z+\tfrac{d-2}4\nabla_zA\ .
}
Hence, the zero-energy wavefunction, annihilated by $Q$, is
\EQ{
\hat\psi_0(z)=\exp\big[-\tfrac{d-2}4A(z)\big]\ .
\label{zes}
}
Notice that such a wavefunction always exists since \eqref{dddg}
always admits the solution where $h_{ab}=h_{ab}(x)$, only, and $\square_x h_{ab}(x)=0$. Furthermore, since the ``Hamiltonian'' $Q^\dagger\cdot Q$ is a 
positive definite Hermitian operator, there are no normalizable negative
energy graviton modes, as required for stability of the gravitational
background.

The condition for having localized four-dimensional gravity is that 
$\hat\psi_0(z)$ is normalizable; in other words
\EQ{
\int d^nz\,\exp\big[-\tfrac{d-2}2A(z)\big]<\infty\ .
}
Notice that normalizability of the ground state
wavefunction is equivalent to the condition that the four-dimensional
gravitational coupling (via \eqref{4dplanck}) be non-vanishing; indeed
\EQ{
G_N \sim \frac{M^{2-d}_*\hat\psi_0(0)^2}{\langle\hat\psi_0|\hat\psi_0\rangle}\
.
}
This requires that $A(z)\to\infty$ sufficiently fast as
$|z|\to\infty$. Normalizability is intimately connected with the
asymptotic behaviour of the potential of the Schr\"odinger equation
\eqref{mdschreq}. If $V(z)>0$ as $|z|\to\infty$, then $\hat\psi_0(z)$
is always normalizable. On the contrary, if $V(z)<0$ as $|z|\to\infty$, then 
$\hat\psi_0(z)$ is not normalizable and therefore is of no interest to us
since it cannot describe localized four-dimensional gravity. The
situation where $V(z)=0$ as
$|z|\to\infty$, is, perhaps, the most interesting and we will focus on
that case in most of the remainder of this paper. 

At this point, we make the obvious remark that in any scenario where
the transverse space is asymptotically flat Euclidean space ($A(z)\to$
constant, as $|z|\to\infty$) $\hat\psi_0(z)$ is non-normalizable and
gravity cannot be localized.

\subsection{Corrections to Newton's Law}\label{sec:newtonslaw}

In order to have localized four-dimensional gravity, we also require
that the other solutions of the Schr\"odinger equation
\eqref{mdschreq}, the KK modes,
do not lead to unacceptably large corrections to
Newton's Law in the four-dimensional theory.

In any realistic brane scenario, the matter fields in the
four-dimensional theory on the brane would be
smeared over the width of the brane in the transverse space. Rather
than deal with this complication, we will, 
for simplicity, consider the gravitational
potential between two point-like sources of mass $M_1$ and $M_2$
located at the origin, $z^i=0$, in the transverse space \cite{RS,ADDK,Kostas}. 
This assumption is justified in cases when the thickness of
the brane is small compared with the bulk curvature.
We expect that
our conclusions will be at least qualitatively correct 
in a general case and present some supporting arguments in Sec.~\ref{sec:con}.
In order to evaluate the correction to Newton's Law, 
we note that a discrete eigenfunction
(these are not present in the RS case) 
of the Schr\"odinger equation $\psi_m(z)$ of energy $m^2$ acts
in four-dimensions like a field of mass $m$ and consequently
contributes a Yukawa-like correction to the 
four-dimensional gravitational potential between two masses $M_1$
and $M_2$:
\EQ{
U(r)\thicksim G_N\frac{M_1M_2}r+
M^{2-d}_*\frac{M_1M_2e^{-mr}}r\psi_m(0)^2
\label{exps}
}
where the wavefunction $\psi_m(z)$ is normalized $\int
d^nz\,\psi_m(z)^2=1$.\footnote{Note that in our convention the
zero-energy state $\hat\psi_0(z)$ is {\em not\/} unit normalized;
however, since we have chosen $A(0)=0$ we have $\hat\psi_0(0)=1$.}
As long as $m$ is large enough, this will be a small correction. The
fact that $\psi_m(0)$ appears in \eqref{exps} is due to our simplifying
assumption that the sources are point-like and located at $z^i=0$. A
more complete analysis would involve the effects of the overlap of the
gravitational modes with the matter modes, and would correct the factors
of $\psi_m(0)$.  We will not have more to say about such corrections here.

The correction from any continuum states $\psi_m(z)$ is obtained by integrating
over these states with the relevant density-of-states
measure. For states which form a continuum in $n$-dimensions starting
at $m_0$ the correction to Newton's Law is
\EQ{
U(r)\thicksim
G_N\frac{M_1M_2}r+M^{2-d}_*\int_{m_0}^\infty dm\, m^{n-1}\,
\frac{M_1M_2e^{-mr}}{r}\psi_m(0)^2\ ,
\label{cnl}
}
where the wavefunctions $\psi_m(z)$ of the continuum are normalized as
plane waves, {\it i.e.\/}~to unity over a period at
$|z|\to\infty$. The factor of $m^{n-1}$ is just the $n$-dimensional plane
wave continuum density of states (up to a constant angular factor).  Notice 
that in flat $d$-dimensional space there would
be no normalizable zero-energy wavefunction $\hat\psi_0(z)$ 
and the continuum would extend down to
$m=0$ and be unsuppressed: $\psi_m(0)=1$. In such a  case $U(r)\sim
M_1M_2M^{2-d}_*r^{-n-1}=M_1M_2M^{2-d}_*r^{3-d}$, as expected for the gravitational
potential in $d$-dimensions.

In the case of intersecting
branes, there are continuum modes which are localized in a
subset of the transverse dimensions. In other words, the wavefunctions behave
as plane waves only in $p<n$ of the transverse dimensions. In this case the
corrections from this part of the continuum spectrum are of the form \eqref{cnl}
with $n$ replaced by $p$.

For the case when
$V(z)>0$ as $|z|\to\infty$, the excited states are clearly separated by a
gap from the ground-state. Hence corrections to Newton's Law are
exponentially suppressed as in \eqref{exps}. As we have already
mentioned, the most interesting case is 
where the potential goes to zero at
infinity. In this case there is a continuum of scattering states
$\psi_m(z)$ with eigenvalues $m^2\geq0$ and so the behaviour of the soft
modes at $z^i=0$ is crucial for determining whether the
corrections are small. 

\subsection{Extension to non-conformally-flat backgrounds}\label{sec:gen}

In this section, we briefly indicate how the preceding analysis of
gravitational fluctuations 
extends to cases where the background is of the general
non-conformally flat form
\eqref{metric}. In order to derive an equation for metric
fluctuations 
\begin{equation}
ds^2=e^{-A(z)}(\eta_{ab}+h_{ab}(x,z))\,dx^a\,dx^b-g_{ij}(z)dz^i\,dz^j\ ,
\label{ourm}\end{equation}
we follow essentially the same steps as for the conformally-flat case
detailed in Sec.~\ref{sec:fluct}. First of all, it is convenient 
to define the metric $\tilde
g_{\mu\nu}(z)=e^{A(z)}g_{\mu\nu}(z)$ and apply the relation \eqref{conformal} in
order to find the variation of the Einstein tensor
$G_{\mu\nu}$. Rather than writing down all of the terms, as we did in
Sec.~\ref{sec:fluct}, we will make immediate use of the following four facts:
(i) $\partial_\mu h^{\mu\nu}=0$; (ii) $h^\mu_\mu=0$; (iii)
$g_{\mu\nu}$ only depends on $z$; and (iv) the only non-vanishing
components of the variation $h_{\mu\nu}$ are $h_{ab}$. By brute force
one can show that the variation of the Einstein tensor
$\widetilde{G}_{\mu\nu}$ is
\SP{
\delta\widetilde{G}_{\mu\nu}=-\tfrac12\tilde\nabla^\rho\tilde\nabla_\rho h_{\mu\nu}=
-\tfrac12\partial^c\partial_ch_{\mu\nu}
-\tfrac12\tilde g^{-1/2}\partial_i(\sqrt{\tilde
g}\tilde g^{ij}\partial_jh_{\mu\nu})\ ,
}
where, until further notice, indices are raised and lowered with
$\tilde g_{\mu\nu}$.
Using 
\eqref{conformal}, we can then write down the variation of the
original Einstein tensor:
\SP{
\delta{G}_{\mu\nu}&=-\tfrac12\partial^c\partial_ch_{\mu\nu}
-\tfrac12\tilde g^{-1/2}\partial_i(\sqrt{\tilde
g}\tilde g^{ij}\partial_jh_{\mu\nu})
+\tfrac{d-2}4\partial_iA\partial^ih_{\mu\nu}\\
&\qquad\qquad\qquad\qquad+\tfrac{d-2}{2}\big(-\tilde g^{-1/2}\partial_i(\sqrt{\tilde
g}\tilde g^{ij}\partial_jA)+\tfrac{d-3}4\partial_iA\partial^iA\big)h_{\mu\nu}\
.
}
Assuming that the variation of the stress-tensor is given by
\eqref{tst}, we have
\EQ{
\delta T_{\mu\nu}=\tfrac{d-2}{2}\kappa^{-2}\big(-\tilde g^{-1/2}\partial_i(\sqrt{\tilde
g}\tilde g^{ij}\partial_jA)+\tfrac{d-3}4\kappa^{-2}\partial_iA\partial^iA\big)h_{\mu\nu}\ .
}
Hence, Einstein's equation gives
\EQ{
-\tfrac12\partial^c\partial_ch_{\mu\nu}
-\tfrac12\tilde g^{-1/2}\partial_i(\sqrt{\tilde
g}\tilde g^{ij}\partial_jh_{\mu\nu})+\tfrac{d-2}4\partial_iA\partial^ih_{\mu\nu}=0\
,
}
which can be rewritten in terms of the original metric as
\EQ{
\frac1{\sqrt{g}}\partial_\rho\big(\sqrt{g}g^{\rho\kappa}\,\partial_\kappa
h_{\mu\nu}\big)=0\ ,
\label{genfluct}
}
where now indices are raised and lowered with $g_{\mu\nu}$. In other
words, the general equation for the fluctuations is simply the
covariant scalar wave-equation. This has been noted in the case of one
transverse dimension in \cite{Kostas} which also discusses its
significance within the AdS/CFT
correspondence. For a conformally-flat background
\eqref{genfluct} reduces to \eqref{dddg}.

For a fluctuation of the form 
$h_{ab}(x,z)=\varphi(z)\check h_{ab}(x)$, with $\square_x\check
h_{ab}(x)=m^2\check h_{ab}(x)$, we have
\EQ{
-\frac1{\sqrt{g}}\partial_i\big(\sqrt{g}g^{ij}\,\partial_j\varphi(z)\big)=m^2g^{00}\varphi(z)\
.
\label{mgenf}
}
This describes the $z$-dependence of a
mode with four-dimensional Kaluza-Klein mass $m$. Notice that
when $m=0$ \eqref{mgenf} always admits the solution $\varphi(z)=$ constant;
this will lead to the analogue of the zero-energy solution
$\hat\psi_0(z)$ in the conformally-flat case. The fluctuation equation
\eqref{genfluct} is derivable from the action
\EQ{
S \thicksim \int d^dx\,\sqrt{g}\partial_\rho
h_{\mu\nu}\partial^\rho h^{\mu\nu}=
\int d^nz\,g^{00}(z)\sqrt{g(z)}\varphi(z)^2\cdot\int
d^4x\,\partial_c\check h_{ab}(x)\partial^c\check h^{ab}(x)+\cdots\ .
\label{gens}
}
{}From the $z$-integral above, inserting $\varphi(z)=$ constant, 
we deduce the generalized expression for the
zero-energy ``wavefunction'' $\hat\psi_0(z)$:
\EQ{
\hat\psi_0(z)=\big[g^{00}(z)\sqrt{g(z)}\,\big]^{1/2}\ ,
\label{gzes}
}
which reduces to \eqref{zes} in the conformally-flat case. The
normalizability condition is consequently 
\EQ{
\int d^nz\,g^{00}(z)\sqrt{g(z)}<\infty\ .
}
The ``wavefunction'' is consequently
\EQ{
\psi(z)\equiv\varphi(z)\big[g^{00}(z)\sqrt{g(z)}\,\big]^{1/2}\ ,
} 
which satisfies a
generalization of the Schr\"odinger equation \eqref{mdschreq}:
\EQ{
-\frac1{[g^{00}\sqrt{g}\,]^{1/2}}
\partial_i\Big(\sqrt{g}g^{ij}\,\partial_j\frac{\psi(z)}{[g^{00}\sqrt{g}\,]^{1/2}}
\Big)=m^2\psi(z)\
.
\label{gseq}
}

\section{Gravity Localized on Thick Three-Branes}\label{sec:domain}

In this section we consider in some detail the case when the 3-brane
is embedded in a space with dimension $d\geq5$. We will, as per
Sec.~\eqref{confflat}, restrict ourselves mainly to a conformally-flat
background which is, in addition, radially symmetric in the space transverse to
the brane. In other words we shall consider metrics of the form
\EQ{
ds^2=e^{-A(\varrho)}\big(\eta_{ab}\,dx^a\,dx^b-d\varrho^2-\varrho^2d\Omega^2\big)
\ ,
\label{radmet}
}
where $\varrho$ is the radial coordinate in the transverse directions and
$\Omega$ are the angular coordinates on $S^{d-5}$. The function
$A(\varrho)$, 
as indicated, only depends on the radial variable. 
We will briefly consider the more general radially symmetric
background which is not necessarily conformally flat at the end of Sec.~\ref{sec:dgf}.

\subsection{Localization and decoupling in $d=5$}\label{sec:locdec}

We begin by discussing the case in $d=5$, corresponding to the RS scenario,
when the transverse space is one-dimensional. In this case, the metric
\eqref{radmet} is
\EQ{
ds^2=e^{-A(z)}\big(\eta_{ab}\,dx^a\,dx^b-dz^2\big)\ .
\label{metfive}
}
The Schr\"odinger equation \eqref{mdschreq} is:
\EQ{
-\frac{d^2\psi(z)}{dz^2}+\big[\tfrac9{16}A'(z)^2-\tfrac34A''(z)\big]
\psi(z)=m^2\psi(z)\
.
\label{scheq}
}
The zero-energy state \eqref{zes} is \cite{Chamblin,Gremm}
\EQ{
\hat\psi_0(z)=\exp\big[-\tfrac34A(z)\big]\ .
\label{zesf}
}
For this to be normalizable, $\exp[-\tfrac32A(z)]$ must fall off faster
than $1/z$. 
As we explained in Sec.~\ref{sec:fdgrav}, the question of whether $\hat\psi_0(z)$ is 
normalizable is intimately connected with the
asymptotic behaviour of the potential $V(z)$. 
If $V(z)>0$ as $|z|\to\infty$, then $\hat\psi_0(z)$
is always normalizable. On the contrary, if $V(z)<0$ as $|z|\to\infty$, then 
$\hat\psi_0(z)$ is not normalizable and therefore is no interest to us
since it cannot describe localized four-dimensional gravity. 

The fact that there are no bound-states with 
negative energy follows from the factorization \eqref{susy}:
\EQ{
\Big[-\frac{d}{dz}+\frac34A'(z)\Big]\Big[\frac{d}{dz}+\frac34A'(z)\Big]
\psi(z)=m^2\psi(z)\ ,
\label{susyf}
}
which has the form of supersymmetric quantum mechanics $Q^\dagger
Q\psi(z)=m^2\psi(z)$, with $Q\equiv d/dz+\tfrac34A^\prime(z)$. Hence the
zero-energy state \eqref{zesf}, which satisfies the supersymmetric
condition $Q\hat\psi_0(z)=0$, is the ground state, the bound-state of
lowest energy. 

\begin{figure}
\begin{center}
\leavevmode \epsfxsize=7cm \epsfbox{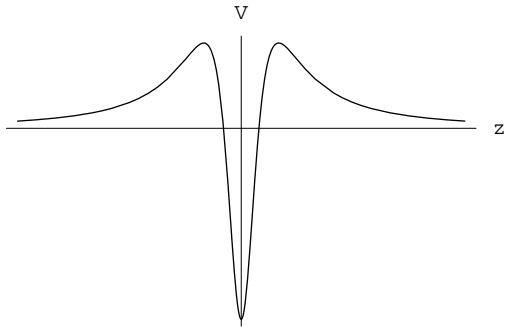}
\end{center}
\caption{\small The Schr\"odinger potential 
\eqref{schpot} for the case $A(z)=\log(k^2z^2+1)$.}
\label{fig:potential}
\end{figure}

The borderline case, $V(z)\to0$ as $|z|\to\infty$, is the most
interesting case, and includes smoothed versions of the 
AdS scenario of \cite{RS}. For
example, when the wavefunction \eqref{zes} has a power-law fall-off
$\psi(z)\sim|z|^{-\alpha}$, then normalizability requires 
$\alpha>\tfrac12$. In this
case the potential $V(z)$ falls off as $\alpha(\alpha+1)/z^2$. 
Conversely, if we assume
that the potential falls off as $|z|^{-2\beta}$ the
wavefunction $\hat\psi_0(z)$
falls off as $e^{-c/|z|^{\beta-1}}$. Consequently,
if $\beta<1$, then the
wavefunction is not normalizable.  It is interesting to note that the
borderline case, where the potential falls off just slowly enough to give
rise to a normalizable bound-state, 
{\it i.e.\/}~$V(z)\sim|z|^{-2}$,
includes the AdS scenario. Figure~\ref{fig:potential} shows the
potential $V(z)$ for a particular smoothing of the AdS
 case with $A(z)=\log(k^2z^2+1)$, 
discussed in Sec.~\ref{sec:ex}.

We now have to consider the question of whether or not the other modes
in our problem decouple. The relevant effects of these modes
on Newton's Law were established in
Sec.~\ref{sec:newtonslaw}. 
For the case when
$V(z)>0$, for $|z|\to\infty$, the excited states are clearly separated by a
gap from the ground-state. Hence corrections to Newton's Law will be
exponentially suppressed, as in \eqref{exps}. The most interesting case is 
where the potential goes to zero at
infinity. In this case there is a continuum of scattering states
$\psi_m(z)$ with eigenvalues $m^2\geq0$. Since the bottom of the
continuum is at $m=0$ it is clear that decoupling is a delicate
issue. From \eqref{cnl}, given that gravity is localized so that 
$\hat\psi_0(z)$ is normalizable, decoupling would require 
that $\int_0^\infty dm\, e^{-mr}|\psi_m(0)|^2$ has no singularity at the
bottom limit of integration.

Before we attempt some rigorous analysis, let us consider the
problem with some rather crude apparatus.
If the continuum modes are to decouple we want the probability
for continuum modes to tunnel into the central region of the potential
to be vanishingly small for $m\to0$. In order
to get a feeling for what might be required, it is instructive to
consider the WKB approximation for this tunneling probability. Consider a
continuum mode with energy $m^2$ incident upon the potential from the
right. The transition
probability is, in the WKB approximation,\footnote{Of course, 
the WKB approximation is not valid in the central region of the potential for
states of small energy. Later we shall present a more rigorous
analysis.}
\EQ{
T(m)\thicksim\exp\Big[-2\int_{z_0(m)}^{z_1(m)}dz\,\sqrt{V(z)-m^2}\Big]\ ,
\label{wkb}
}
where $0<z_0(m)<z_1(m)$ are the two points in the rightmost barrier region
where $V(z)=m^2$. In order for
the soft continuum states to decouple we would need $T(0)=0$. Since
$z_1(0)=\infty$, this can be achieved if the integral \eqref{wkb} has
a divergence for large $z$; in other words, if $V(z)$ falls off at
least as slowly as $1/z^2$. This matches precisely the condition on
the normalizability of the state $\hat\psi_0(z)$. This suggests that
there is a natural decoupling of the continuum states precisely when
the ground state $\hat\psi_0(z)$ is normalizable. 

So the hypothesis that we want to establish is that the continuum
modes decouple (that is lead to small corrections to Newton's
Law) precisely when $\hat\psi_0(z)$ is normalizable. To this end, consider a
potential of the form
\EQ{
V(z)\sim\frac{\alpha(\alpha+1)}{z^2}\ ,
\label{potas}
}
for large $|z|$.
We shall find that the crossover from localization to
de-localization occurs for some critical value of $\alpha$. 
We shall not assume any particular form for the potential $V(z)$
except, for simplicity, that it only depends on a single dimensionful scale $k$, so
that, for instance, the central region extends over a scale $1/k$; in other
words \eqref{potas} is valid for $|z|\gg1/k$.

As we have discussed above,
what we need to calculate in order to investigate the decoupling of
the continuum modes is the limiting behaviour of 
$\psi_m(0)$ at small $m$.\footnote{Recall from Sec.~\ref{sec:newtonslaw} that the
wavefunctions $\psi_m(z)$ must be normalized as plane waves for
$|z|\to\infty$.} 
Let us consider four different regions in $z$ (we will always be
considering modes with energies $m\ll k$): (1) $z\ll 1/k$, (2) $1/k\ll z\ll1/m$, 
(3) $1/k\ll z\sim1/m$ and (4) $1/m\ll z$. 
In regions (2),(3) and (4), we must solve the Schr\"odinger equation with the tail
of the potential \eqref{potas}:
\EQ{
-\frac{d^2\psi_m(z)}{dz^2}+\frac{\alpha(\alpha+1)}{z^2}
\psi_m(z)=m^2\psi_m(z)\
.
}
The solution is given by a linear combination of Bessel functions
\EQ{
\psi_m(z)=a_mz^{1/2}Y_{\alpha+1/2}(mz)+b_mz^{1/2}J_{\alpha+1/2}(mz)\
.
\label{KKwf}
}
In region (4), where $mz\gg1$, the Bessel functions become plane waves:
\EQ{
\psi_m(z)=a_m\sqrt{\frac{2}{\pi m}}\sin(mz-\tfrac\pi2\alpha-\tfrac\pi2)+
b_m\sqrt{\frac{2}{\pi m}}\cos(mz-\tfrac\pi2\alpha-\tfrac\pi2)\ .
\label{regfour}
}
In region (2), where $mz\ll1$ (but $kz\gg1$), the Bessel functions can be expanded in
$mz$ giving\footnote{The corrections in the second square bracket,
coming from the expansion of $J_{\alpha+1/2}(mz)$ for small $mz$, are a
power series in $(mz)^2$. The corrections in the first square bracket,
coming from the expansion of $Y_{\alpha+1/2}(mz)$ are more complicated
since they depend on whether 
$\alpha+1/2$ is an integer, or not.
However, the two terms indicated are the dominant terms for small $mz$.}
\SP{
\psi_m(z)
=-\frac{a_mz^{1/2}\Gamma(\alpha+1/2)}\pi
\left(\frac2{mz}\right)^{\alpha+1/2}&\Big[1+\frac1{\alpha-1/2}
\left(\frac{mz}2\right)^2+\cdots\Big]\\
&+\frac{b_mz^{1/2}}{\Gamma(\alpha+3/2)}
\left(\frac{mz}2\right)^{\alpha+1/2}\Big[1+\cdots\Big]\ .
\label{regt}
}
In other words we can match the wavefunction in regions (2) and (4) by using
the asymptotic behavior of the Bessel functions, and the exact form of the
Bessel function then determines the behavior in the intermediate region, (3).

Now we show how to match regions (1) and (2). In these
regions $mz\ll1$ and it is meaningful to solve the Sch\"odinger equation as a
series in $m^2$. The first two terms are
\EQ{
\psi_m(z)=\hat\psi_0(z)+m^2\phi(z)+\cdots\ ,
\label{expan}
}
where $\hat\psi_0(z)$ is the suitably normalized zero energy solution \eqref{zes}. The
first correction $\phi(z)$ satisfies the inhomogeneous equation
\EQ{
\Big[-\frac{d^2}{dz^2}+V(z)\Big]\phi(z)=\hat\psi_0(z)\ .
\label{inhomo}
}
Now we can match \eqref{expan} with $z\gg1/k$,
with the series in region (2)
\eqref{regt}. In this region $\hat\psi_0(z)\sim z^{-\alpha}$, which matches
the first term in \eqref{regt} if $a_m\sim m^{\alpha+1/2}$. The second
and third terms in \eqref{regt} then match the next term in the expansion
\eqref{expan} as long as $b_m\sim m^{-\alpha+3/2}$.\footnote{A
caveat is that the third term in \eqref{regt}, coming from
$J_{\alpha+1/2}(mz)$, could match terms higher in the expansion \eqref{expan};
however, this would require a non-generic behaviour of the
potential.} 
Using this matching
procedure, we have determined the $m$-dependence of $a_m$ and $b_m$. 
Since $m$ is small, the dominant term in region (4) comes from the
second term in \eqref{regfour} where the coefficient of the cosine
goes like $b_mm^{-1/2}\sim m^{-\alpha+1}$. Hence to normalize the
wavefunction $\psi_m(z)$
as a plane wave, we must multiple it by an overall factor of
$m^{\alpha-1}$. The value of the wavefunction at $z=0$ is then
extracted from \eqref{expan}. To leading order in $m$,
\EQ{
\psi_m(0)\thicksim \left(\frac mk\right)^{\alpha-1}\ ,
\label{smb}
}
where the factor of $k$ is dictated by dimensional analysis. Notice
that in the AdS scenario of \cite{RS}, $\alpha=\tfrac32$, in which
case we find $\psi_m(0)^2\sim m/k$, in agreement with
the exactly solved delta-function potential of \cite{RS}.

Equation \eqref{smb} establishes the asymptotic behaviour of
the continuum modes at $z\ll 1/k$, where $k$ is
characteristic decay width of the potential $V(z)$. If the
matter is smeared over the distances $1/k$ one needs to obtain more
information about the wave-functions of the excited modes to
study the corrections to the Newton's Law. It is important
to note, however, that we obtained \eqref{smb} without
specifying details of the potential near $z=0$. 
Thus, we can easily introduce a potential such that matter
is localized near $z=0$, yet the arguments leading to
\eqref{smb} remain valid.

{}From \eqref{smb}, we find that the integral 
over the continuum modes in \eqref{cnl} is only non-singular at the
bottom limit of integration if $\alpha>\tfrac12$. In this case the
corrections are  
\EQ{
U(r)\thicksim
G_N\frac{M_1M_2}r+\frac{C}{M^3_*k^{2\alpha-2}}
\frac{M_1M_2}{r^{2\alpha}}=G_N\frac{M_1M_2}{r}\Big(1+\frac{C'}{(kr)^{2\alpha-1}}\Big)\ ,
\label{wwsi}
}
where $C$  and $C'$ are dimensionless numbers. We are assuming that
the metric only depends on one dimensionful parameter $k$, so that
$G_N\sim kM^{-3}_*$. Notice that when the potential
falls off as \eqref{potas} there are power-law corrections to
Newton's Law with a universal exponent
$\alpha$ determined simply by the long-range
fall-off of the potential. We emphasize that the AdS case of \cite{RS}
corresponds to taking $\alpha=\tfrac32$, which agrees with the exact
calculation of the correction presented in \cite{RS}.

Now we see that it is precisely when $\hat\psi_0(z)$ is normalizable, {\it
i.e.\/}~when $\alpha>\tfrac12$, that the
continuum states give a correction to Newton's Law which is suppressed
relative to the leading term.

\subsection{Examples}\label{sec:ex}

At this point it is probably worthwhile to consider some 
examples. To begin with, 
consider the class of conformally-flat five dimensional backgrounds for which 
\EQ{
A(z)=\tfrac{2\alpha}3\log(k^2z^2+1)\ ,
\label{exama}
}
for some constants $\alpha$. The case when $\alpha=\tfrac32$ is
particularly interesting because in this case we can easily transform
back to the $r$ coordinate in which case $A(r)=2\log\,\cosh(kr)$. So
when $\alpha=\tfrac32$ the space given by \eqref{exama} is asymptotically AdS
and therefore represents a smoothing of the AdS example of
\cite{RS}, recently discussed in \cite{Gremm}. 

For general $\alpha$, the potential in the Schr\"odinger
equation and the ground-state
corresponding to \eqref{exama} are, respectively,
\EQ{
V(z)={k^2\alpha}\frac{(\alpha+1)k^2z^2-1}{(k^2z^2+1)^2}\
,\qquad\hat\psi_0(z)=\frac1{(k^2z^2+1)^{\alpha/2}}\ .
\label{expotgs}
}
In other words, this class of examples has precisely the asymptotic
form of the potential discussed in the last section. The shape of the
potential for  $\alpha=\tfrac32$  appears in Figure \ref{fig:potential}.

\begin{figure}
\begin{center}
\leavevmode \epsfxsize=7cm \epsfbox{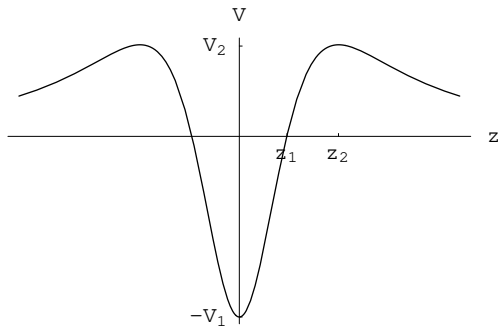}
\end{center}
\caption{\small The shape of the potential \eqref{V2}.}
\label{fig:shape}
\end{figure}
The example above has the advantage of being simple;
however, 
since it only depends on one parameter, $k$, the thickness
of the brane is comparable to the bulk curvature. Without
any further assumptions, one expects that the matter fields
living on the brane will be smeared over the distance $1/k$
and such smearing may affect the estimates for the
corrections to the Newton's law. In order to localize the
matter fields near $z=0$ we can introduce a
parameter characterizing thickness of the brane (as observed
by the matter fields) in addition
to the overall scale $k$.
For example, in the RS scenario, we could replace $|r|$ in
$A(r)=2k|r|$ by either
$r\,{\rm tanh}(\mu r)$ or $\mu^{-1}\log\cosh(\mu r)$, which
both depend on a thickness paramater $\mu$;
however, in these cases we cannot calculate $z=z(r)$ and hence $V(z)$ in
closed form. It is consequently more convenient to start in the
$z$-coordinate basis with $A(z)=\tfrac{4\alpha}3\log(k|z|+1)$ (where the RS
scenario has $\alpha=\tfrac32$) and smooth $|z|$ with either
$z\,{\rm tanh}(\mu z)$ or $\mu^{-1}\log\cosh(\mu z)$. In the latter case
\EQ{
A(z)=\tfrac{4\alpha}3\log(k\mu^{-1}\log\cosh(\mu z)+1)\ ,
\label{gexama}
}
from which we can easily calculate the potential 
\EQ{
V(z)=\frac{\alpha k\mu^2}2\cdot\frac{k(\alpha+1)(\cosh(2\mu
z)-1)-2\mu-2k\log\cosh(\mu z)}{\cosh^2(\mu z)\big(k\log\cosh(\mu
z)+\mu\big)^2}\ , 
\label{V2}
}
which has the asymptotic form \eqref{potas} for $|z|\gg{\rm max}(\mu^{-1},k^{-1})$.
The generic shape of this potential is illustrated in
Figure~\ref{fig:shape} where we have introduced the parameters
$V_1=-V(0)$,  $V_2={\rm max}[V(z)]$ and $z_1$, defined by $V(z_1)=0$.
The depth of the central well is easily found to be $V_1=\alpha k\mu$. 
In the limit $\mu\gg k$ the brane is
very thin and, for $\alpha=\tfrac32$, the potential approaches that of Randall and Sundrum
\cite{RS}. In this limit for general $\alpha$, $V_2\sim k^2$, independent
of $\mu$, while $z_1$ is asymptotically
\EQ{
z_1=\frac1{2\mu}\log\frac{4\mu}{(\alpha+1)k}\ ,\qquad \mu\gg k\ ,
}
so the central well becomes more
like a delta function as $\mu\gg k$.
In this limit $k$ controls the long-range fall-off of the potential,
via the asymptotic form
\EQ{
V(z)=\frac{\alpha(\alpha+1)k^2}{(k|z|+1)^2}\ ,\qquad |z|\gg\mu^{-1}\ .
}
So in the limit $\mu\gg k$ we would expect that
matter fields are localized near $z=0$, in which case we can neglect the 
overlap of those fields with the gravity modes as corrections to
Newton's law and the formulae of Sec.~\ref{sec:newtonslaw} will be valid.
The other limit $\mu\ll k$, where 
the brane is much thicker than $k$, is also interesting. In this
limit, both $V_1$ and $V_2$ scale like $k\mu$, while $z_1$ asymptotes to
\EQ{
z_1=\sqrt{\frac2{2\alpha+1}}(k\mu)^{-1/2}\ ,\qquad\mu\ll k\ .
}
In this limit we expect that the smearing of the matter fields
over the transverse direction will become important and 
that the analysis of the corrections to Newton's Law as described in
Sec.~\ref{sec:newtonslaw} will require some modification.

\subsection{Resonant modes}\label{sec:res}

The final issue that we mention regarding decoupling, is the possible
existence of resonances. It can happen that for some particular
energies, incident plane waves can resonate with the potential $V(z)$
and consequently have a large value of $\psi_m(0)$.  This possibility was
also noticed in \cite{Gremm}.

\begin{figure}
\begin{center}
\leavevmode \epsfxsize=7cm \epsfbox{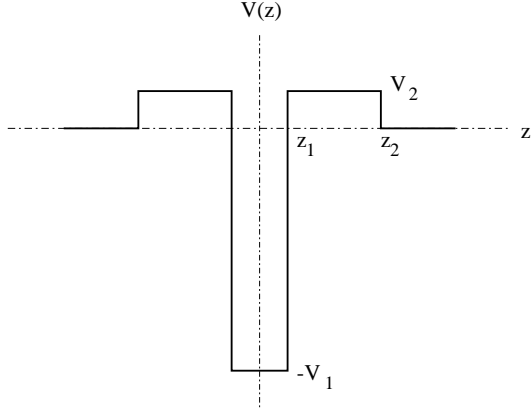}
\end{center}
\caption{\small The volcano box potential.}
\label{fig:volbox}
\end{figure}

A useful toy model in which the Kaluza-Klein modes can be calculated
exactly is the volcano box potential (Figure~\ref{fig:volbox}).  Because
the potential is exactly zero beyond the barrier the can be no bound-state
at zero energy, but the depth and width of the well can be easily arranged
such that there is a single bound-state, with a vanishing small energy
$m^2<0$, and a continuum for $m^2\geq0$.
The solution for a symmetric continuum wavefunction (orbifold
boundary conditions, as in \cite{RS2}) is: 
\EQ{
\psi(z)=\begin{cases} \cos k_1z & |z|\leq z_1\\
a\,e^{k_2z}+b\,e^{-k_2z} & z_1\leq |z|\leq z_2\\
 c\,\cos k_3x +d\,\sin k_3x & |z|\geq z_2\ ,\end{cases}
}
where 
\EQ{
k_1=\sqrt{m^2+V_1}\ ,\qquad
k_2=\sqrt{V_2-m^2}\ ,\qquad
k_3=\sqrt{m^2}\ ,
}
and the coefficients are given by 
\AL{
a&=\tfrac{e^{-k_2z_1}}{2}\big(\cos k_1z_1-\tfrac{k_1}{k_2} \sin k_1z_1
\big)\ ,\qquad
b=\tfrac{e^{k_2z_1}}{2}\big(\cos k_1z_1+\tfrac{k_1}{k_2} \sin k_1z_1
\big)\ , \\
\begin{split}
c&=\tfrac{e^{-k_2(z_1+z_2)}}{2k_2k_3}\big[k_2 \cos k_1z_1\big(
(e^{2k_2z_1}+e^{2k_2z_2})k_3 \cos k_3z_2 +(e^{2k_2z_1}-e^{2k_2z_2})k_2 
\sin k_3z_2\big)\\
&+k_1 \sin k_1z_1\big((e^{2k_2z_1}-e^{2k_2z_2})k_3
 \cos k_3z_2+(e^{2k_2z_1}+e^{2k_2z_2})k_2 \sin k_3z_2\big)\big]\ , 
\end{split}\\
\begin{split}
d&=\tfrac{e^{-k_2(z_1+z_2)}}{2k_2k_3}\big[k_2 \sin k_1z_1\big(
(e^{2k_2z_1}+e^{2k_2z_2})k_2 \cos k_3z_2
-(e^{2k_2z_1}-e^{2k_2z_2})k_3 
\sin k_3z_2\big)\\&+k_2 \cos k_1z_1\big(-(e^{2k_2z_1}-e^{2k_2z_2})k_2
 \cos k_3z_2+(e^{2k_2z_1}+e^{2k_2z_2})k_3 \sin k_3z_2\big)\big]\ .
\end{split}
}
Resonant modes occur when the coefficient $a$ of the growing exponential
in the region $z_1\leq|z|\leq z_2$,
vanishes, {\em i.e.} 
\EQ{
\cos\,k_1z_1-\frac{k_1}{k_2}\,\sin\,k_1z_1 =0\  . 
\label{a=0}
}
We assume that $m^2<V_2\ll V_1$.  In order to make contact with smooth
versions
of the RS model we set $V_1z_1=k$ and $V_2=k^2$, up to numerical
coefficients.
The smoothings can in general introduce other dimensionful quantities
besides $k$; for instance,
the width $2z_1$, or equivalently the depth $V_1$, of the well part of the 
potential.  If we in addition take $(V_1z_1)^2\ll V_1$ 
(or $x_1\sqrt{V_2}\ll 1$) then we can expand 
\eqref{a=0} in powers of $(V_1z_1)^2/V_1$ and $m^2/V_1$ to obtain, 
\EQ{
1-\frac{(V_1z_1)^2}{2V_1}\left(1+\frac{m^2}{V_1}\right)-\frac{V_1z_1
}{\sqrt{V_2-m^2}}\Big(1+\frac{m^2}{V_1}\Big) 
\thicksim 1-\frac{V_1z_1}{\sqrt{V_2-m^2}}=0 \ . \label{Vz}
}
If there is a solution, then it is $m^2\sim V_2-(V_1z_1)^2 \sim k^2$.  
The spacing
between resonances would be of order $\pi^2/z_1^2$, so if 
there is a resonance below the barrier height $V_2$
there is only one (for narrow wells).  
The contribution of the narrow
resonance to Newton's law would be of the form
\EQ{
U(r)\thicksim \frac{e^{-mr+z_2\sqrt{V_2-m^2}}}{r}\ . 
}
Hence, the contribution of the resonance is negligible for 
$r\gg z_2\sqrt{V_2-m^2}/m^2$.
Notice that by \eqref{Vz} if  $V_1z_1>\sqrt{V_2}$ there will not
be a resonance at all.  In the RS case the potential is 
\cite{RS},
\EQ{
V(z)=\frac{15k^2}{4(k|z|+1)^2}-3k\delta(z)\ . 
}
If we na\"\i vely read off the coefficients $3k$ and $\tfrac{15}4k^2$ of the delta
function and barrier height, respectively, then we find that there are
no resonant modes in the volcano box approximation.  However, this
matching
of the RS case to the volcano box approximation is too glib and a more
careful analysis is needed.
Hence, we find that if the well is deep, there is at most a single 
resonance at
the order of, but below, the barrier height.  We expect that this will
be the case for general smoothings of the RS scenario, but in the absence
of
explicit solutions for the wavefunctions it is difficult to make definite 
predictions.  In any case, because the resonances give rise to Yukawa type
contributions to the gravitational potential, they are irrelevant below a
certain scale which for smooth versions of the RS case is expected to be
of order the fundamental Planck scale $M_*$.

\begin{figure}
\begin{center}
\leavevmode \epsfxsize=7cm \epsfbox{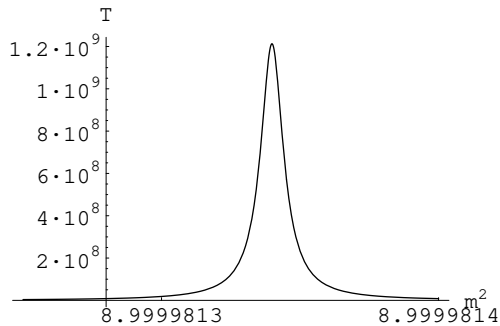}
\end{center}
\caption{\small Resonance of transmission $T$ into well with Kaluza-Klein 
mass $m$.}
\label{fig:res1}
\end{figure}

For example, in Figure~\ref{fig:res1} we plot the 
``transmission coefficient'' $T=1/(c^2+d^2)$ (which is {\it not\/}
restricted to $T\leq1$) into the well
versus the energy of the KK mode
$m^2$.  In this example, in units where the fundamental Planck scale
$M_*=1$, we take $V_1=10^6,V_2=10,z_1=1/V_1$ and $z_2=10$.  We find a sharp
resonance 
near $m^2=V_2-(V_1z_1)^2=9$, as expected, of width $(\Delta m^2)/m^2\sim
10^{-8}$.  Away from the resonance $T$ is smaller than ${\cal O}(10^{-4})$,
except near $V_2$, where $T$ increases to $.012$ as a result of the weakening 
effect of the potential barrier at higher energies.


\subsection{Localization and decoupling in $d>5$}\label{sec:dgf}

In this section, we consider the generalization to the cases when the
transverse space is more than five dimensional. 
In this case, it is convenient to use polar coordinates for
$z=(\varrho,\Omega)$ and 
write $\psi(z)=R(\varrho)Y_l(\Omega)$, where
$Y_l(\Omega)$ is an $n$-dimensional spherical harmonic, with
\EQ{
\nabla_z^2Y_l(\Omega)=-
\frac{l(l+d-6)}{\varrho^2}Y_l(\Omega)\ .
}
{}From \eqref{mdschreq}, we find that the radial function $R(\varrho)$ then satisfies
\SP{
-R^{\prime\prime}(\varrho)-\tfrac{d-5}{\varrho}R'(\varrho)+
&\Big[\tfrac{(d-2)^2}{16}A'(\varrho)^2
-\tfrac{d-2}4A^{\prime\prime}(\varrho)\\
&-\tfrac{(d-2)(d-5)}{4\varrho}A'(\varrho)+\tfrac{l(l+d-6)}{\varrho^2}\Big]R(\varrho)
=m^2R(\varrho)\ .
\label{hdse}
}
The zero-energy state which potentially describes a four-dimensional graviton
is from \eqref{zes}
\EQ{
\hat\psi_0(\varrho)=\exp\big[-\tfrac{d-2}4A(\varrho)\big]\ .
}

We now want to argue that, just as in the case in $d=5$, the
wavefunction $\hat\psi_0(z)$ becomes normalizable precisely when the
KK modes decouple. As in $d=5$, the most delicate case is 
when $\hat\psi_0(z)$ has a power-law fall-off $\sim
\varrho^{-\alpha}$. Normalizability requires that $\alpha>\tfrac{d-4}{2}$.
In this case the potential in \eqref{hdse} 
falls off as $\sim\varrho^{-2}$. It is clear that 
only the $s$-wave modes ($l=0$) are non-vanishing at the
origin. Furthermore, we do not expect the matter fields of interest to us to
transform under the ${\rm SO}(n)$ global symmetry and hence they would only
couple at tree level to the $s$-wave modes. Finally, even for fields
transforming non-trivially under ${\rm SO}(n)$ the sum over $l$ presumably
leads to a convergent series.\footnote{We thank Martin
Gremm for raising the issue of the $l>0$ modes.} For this reason we shall
only consider the $l=0$ modes here. We now follow exactly same steps as
in Sec.~\ref{sec:locdec}. Asymptotically for large $\varrho$, in regions  (2),(3)
and (4), the radial function satisfies the equation, for $l=0$,
\EQ{
-R^{\prime\prime}(\varrho)-\tfrac{d-5}{\varrho}R'(\varrho)+\tfrac{\alpha(\alpha+6-d)}
{\varrho^2}R(\varrho)=m^2R(\varrho)\ .
\label{besseq}
}
As previously, we can solve this equation in terms of Bessel functions: for $l=0$
\EQ{
R(\varrho)=a_m\varrho^{3-d/2}
Y_{\alpha+3-d/2}(m\varrho)+b_m\varrho^{3-d/2}J_{\alpha+3-d/2}(m\varrho)\
.
\label{besssol}
}
We now match the solution to that in region (1) using the same
procedure that we followed in Sec.~\ref{sec:locdec}. In this case, we
find after normalizing the wavefunctions as plane planes at $\varrho\to\infty$,
\EQ{
\psi_m(0)\thicksim \left(\frac mk\right)^{\alpha-d+4}\ ,
\label{bessbe}
}
Plugging this into the correction to Newton's Law, we find that the
integral over the $s$-wave KK modes has no singularity if
$\alpha>\tfrac{d-4}2$, matching the condition that $\hat\psi_0$ is
normalizable, and the correction is of the form
 \EQ{
U(r)\thicksim
G_N\frac{M_1M_2}r+\frac{C}{M^{d-2}_*k^{2\alpha-2d+8}}\frac{M_1M_2}
{r^{2\alpha-d+5}}=
G_N\frac{M_1M_2}r\Big(1+\frac{C'}{(kr)^{2\alpha-d+4}}\Big)\ ,
}
where $C$ and $C'$ are some dimensionless numbers and
we have assumed that the metric depends upon a single
dimensionful parameter $k$ so that $G_N\sim M^{2-d}_*k^{d-4}$.

Although we have only presented a detailed analysis for the conformally-flat
backgrounds, it is a simple matter, using the formulae of
Sec.~\ref{sec:gen}, 
to generalize to the arbitrary
radially symmetric background having the form
\EQ{
ds^2=e^{-A(\varrho)}\eta_{ab}\,dx^a\,dx^b-e^{-B(\varrho)}
\big( d\varrho^2+ \varrho^2d\Omega^2\big)\ ,
\label{gradmet}
}
where the coordinates $\{\varrho,\Omega\}$ are polar coordinates
in $n=(d-4)$-dimensions. In the general case there is no canonical
choice for the $\{z^i\}$ coordinates. It turns out that in order to
analyze the localization of gravity it is {\em not\/} judicious to
choose the $\{z^i\}$ to be the Cartesian coordinates associated to  
$\{\varrho,\Omega\}$. On the contrary, we will choose the  $\{z^i\}$
to be the Cartesian coordinates associated to polar coordinates
$\{\tilde\varrho,\Omega\}$, involving a new radial coordinate
$\tilde\varrho=\tilde\varrho(\varrho)$, for which the metric
\eqref{gradmet} has the form
\EQ{
ds^2=e^{-\tilde A(\tilde\varrho)}\big(\eta_{ab}\,dx^a\,dx^b-
d\tilde\varrho^2\big)-e^{-\tilde B(\tilde\varrho)}
\tilde\varrho^2d\Omega^2\ ,
\label{ggradmet}
}
where the functions $\tilde A(\tilde\varrho)$ and $\tilde
B(\tilde\varrho)$ are related to $A(\varrho)$ and $B(\varrho)$ by the
coordinate transformation on the radial coordinate. In the new set of
coordinates $\{z^i\}$ the wavefunction is \eqref{gzes}
\EQ{
\hat\psi_0(z)=\exp\big[-\tfrac34\tilde
A(\tilde\varrho)-\tfrac{d-5}4\tilde B(\tilde\varrho)\big]
=\exp\big[-\tfrac{d-2}4\hat A(\tilde\varrho)\big]\ ,
\label{gzewf}
}
where we have defined
\EQ{
\hat A(\tilde\varrho)=\tfrac3{d-2}\tilde
A(\tilde\varrho)+\tfrac{d-5}{d-2}\tilde B(\tilde\varrho)\ .
}
To be completely explicit, the normalizability condition is
\EQ{
\int d^nz\,\hat\psi_0(z)^2={\rm Vol}(S^{d-5})\,
\int \tilde\varrho^{n-1}d\tilde\varrho\,\exp\big[-\tfrac{d-2}2\hat 
A(\tilde\varrho)\big]<\infty\ .
}
The equation for the fluctuations \eqref{gseq} can be simplified by
separating the variables: $\psi(z)=R(\tilde\varrho)Y_l(\Omega)$, which leads to a
generalization of the radial equation \eqref{hdse}:
\SP{
-R^{\prime\prime}(\tilde\varrho)-\tfrac{d-5}{\tilde\varrho}R'(\tilde\varrho)+
&\Big[\tfrac{(d-2)^2}{16}\hat A'(\tilde\varrho)^2
-\tfrac{d-2}4\hat A^{\prime\prime}(\tilde\varrho)\\
&-\tfrac{(d-2)(d-5)}{4\tilde\varrho}\hat A'(\tilde\varrho)+
\tfrac{l(l+d-6)e^{\tilde B(\tilde\varrho)-\tilde 
A(\tilde\varrho)}}{\tilde\varrho^2}\Big]R(\tilde\varrho)
=m^2R(\tilde\varrho)\ .
\label{ghdse}
}
When $\tilde B(\tilde\varrho)=\tilde A(\tilde\varrho)$ the metric
\eqref{ggradmet} is conformally-flat and the previous equation for the
radial fluctuations for the conformally-flat case 
\eqref{hdse} is recovered. Our equation
\eqref{ghdse} matches that derived in \cite{CK} for the case when the
transverse space is two dimensional. Notice that for $s$-waves the
equation for the radial function $R(\tilde\varrho)$ \eqref{ghdse} is identical to 
\eqref{hdse} with the replacement $A(\varrho)\to\hat
A(\tilde\varrho)$. Moreover the expression for the zero-energy
wavefunction $\hat\psi_0(z)$ \eqref{gzewf} is identical to that in the
conformally-flat case with the same replacement  $A(\varrho)\to\hat
A(\tilde\varrho)$. Consequently, we can use the same analysis as in
the conformally-flat case to argue that gravity is localized, {\it
i.e.\/}~$\hat\psi_0(z)$ is normalized, when the $s$-wave
continuum modes are decoupled.

\subsection{Thick three-branes from a scalar field}\label{sec:scalars}

In this section, we investigate whether the three-brane scenario that
we have discussed in previous sections can actually be generated by
gravity coupled to a single real scalar field.

The first issue that we must verify is that our ansatz \eqref{tst} for the
behaviour of the stress tensor under gravitational fluctuations is,
in fact,
valid for a scalar field. The fact that it {\em is\/} valid follows from the
dependence of the action of the scalar field on the metric. In
general, the total action of gravity plus the scalar takes the form
\EQ{
S=\int d^dx
\,\sqrt{g}\big[-\kappa^{-2}R+\tfrac12\partial_\mu\phi\partial^\mu\phi-{\cal
V}(\phi)\big]\ .
\label{5dscaction}
}
The stress-tensor for the scalar field is
\begin{equation}
T_{\mu\nu}=\tfrac12\partial_\mu\phi\,\partial_\nu\phi-\tfrac12g_{\mu\nu}\big[\tfrac{1}{2}
\partial_\alpha\phi\,\partial_\beta\phi\,g^{\alpha\beta}-{\cal
V}(\phi)\big]\ .
\label{tmn}
\end{equation}
For metric fluctuations $h_{\mu\nu}$ which only
have a non-vanishing component $h_{ab}$, since $\phi=\phi(z)$ only, we have
\EQ{
h^{\mu\nu}\partial_\nu\phi=0\ .
}
{}From this and the form of the stress tensor \eqref{tmn}, the variation of
the stress tensor under metric fluctuations \eqref{tst} follows 
immediately. In the coupled system, the analysis of fluctuations is naturally more involved.
However, the fluctuations of the scalar are completely decoupled from
the transverse traceless gravitational fluctuations as described in Sec.~\ref{sec:fluct}
and consequently all our previous conclusions regarding the localization of
gravity are still valid.

Next we consider whether the three-branes can
be formed from a single real scalar field. 
The system of scalar fields coupled to gravity was studied in the same
context in \cite{DFGK,Chamblin,skenderis,cvetic}. Later, in
Sec.~\ref{sec:intscalar}, we will will
prove that a single real scalar field cannot produce a three-brane
in $d>5$ dimensions, so for the rest of this section we will take $d=5$
and take the metric in the form \eqref{rsmetric}  and take
$\phi=\phi(r)$ only. This ansatz, 
as we will see, completely determines the form of the scalar potential 
${\cal V}(\phi)$ and the solution $\phi(r)$.

The graviton equation of motion is, as usual, $G_{\mu\nu}=
\kappa^2T_{\mu\nu}$, with $T_{\mu\nu}$ as in \eqref{tmn}. 
Plugging in the ansatz \eqref{rsmetric} and $\phi=\phi(r)$, there are
two independent components of Einstein's equation: 
\AL{
&\kappa^2\phi^\prime(r)^2+2\kappa^2{\cal
V}[\phi(r)]+6A^\prime(r)^2-6A^{\prime\prime}(r)=0\ , \\
&\kappa^2\phi^\prime(r)^2-2\kappa^2{\cal V}[\phi(r)]-6A^\prime(r)^2=0\ .
}
It immediately follows that,
\AL{
\kappa^2{\cal
V}[\phi(r)]&=-3A^\prime(r)^2+\tfrac{3}{2}A^{\prime\prime}(r)\ , 
\label{5dV} \\
\kappa^2\phi^\prime(r)^2&=3A^{\prime\prime}(r)\ . \label{phi}
}
Note that a solution only exists for $\phi(r)$ if $A^{\prime\prime}(r)\geq0$.
The scalar field equation following from the action \eqref{5dscaction}
is 
\begin{equation}
\frac{1}{\sqrt{g}}\partial_\mu\big(\sqrt{g}\,g^{\mu\nu}\,\partial_\nu\phi(r)
\big)+\frac{\partial{\cal V}(\phi)}{\partial\phi}=0\ , 
\end{equation}
or, with our ansatz, \begin{equation}
-\phi^{\prime\prime}(r)+2\,A^\prime(r)\,\phi^\prime(r)+\frac{\partial
{\cal V}(\phi)}{\partial\phi}=0\ .
\end{equation}
One can easily show that the scalar field equation is solved
automatically by a solution of Einstein's equation. To see this we
note that Einstein's equation implies $\nabla_\mu T^{\mu\nu}=0$, which
itself implies the scalar field equation due to the general covariance
of the scalar field action.\footnote{We are grateful to Shanta de Alwis for
pointing this out.} Hence, given the scalar potential 
${\cal V}(r)$, any solution to Einstein's equation 
will automatically be a solution to the scalar field
equations.  Alternatively, any metric of the form \eqref{rsmetric} is
a solution to the gravity and scalar equations-of-motion if the scalar field
has the form \eqref{phi} and the scalar potential is given by 
\eqref{5dV}.

Furthermore, if $\phi(r)$ is a strictly monotonic function of $r$ then we  
can we can implicitly define 
\begin{equation}
{\cal W}[\phi(r)]\equiv\kappa^{-1}\,A^\prime(r)\ ,
\label{BPS1}
\end{equation}
hence it follows from \eqref{phi} that 
\begin{equation}
\phi^\prime(r)=3\kappa^{-1}\frac{\partial {\cal
W}[\phi(r)]}{\partial\phi(r)}\ .
\label{BPS2}
\end{equation}
Then we can write the scalar potential as, 
\begin{equation}
{\cal V}[\phi]=\tfrac{9}{2}\big(\frac{\partial{\cal W}(\phi)}{\partial\phi}
\big)^2-3{\cal W}(\phi)^2\ . \label{5dspot}
\end{equation}
We therefore see that the supersymmetric form of the scalar potential and BPS
equations introduced in \cite{DFGK,Chamblin} appear naturally in this
approach.  The ansatz \eqref{rsmetric} for the metric determines
that the scalar potential can be written in the supersymmetric form
\eqref{5dspot}.

\section{Gravity Localized on Thick Intersecting Branes}

In this section, we consider the possibility that gravity can be
localized on the four-dimensional intersection of higher dimensional
branes. In other words, we search for conformally-flat backgrounds 
\eqref{confflat} with $d>5$, that we can interpret as a
four-dimensional intersection of $d-4$ $(d-1)$-branes.

\begin{figure}
\begin{center}
\leavevmode \epsfxsize=7cm \epsfbox{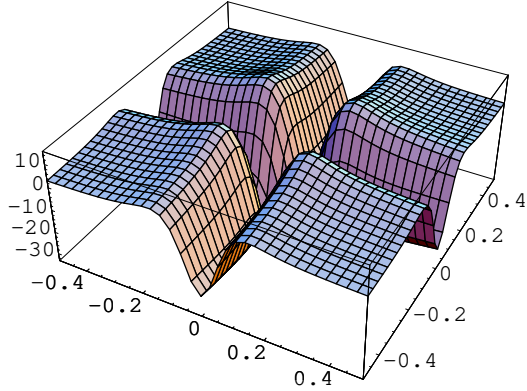}
\end{center}
\caption{\small The Schr\"odinger potential $V(z_1,z_2)$
for the metric \eqref{mdrsmetric} 
with $k=1$ and $|z|\to z\,\tanh(10z)$.}

\label{fig:6d}
\end{figure}

For example, we have in mind smoothings of the multi-dimensional AdS
metric in \eqref{mdrsmetric}. For instance, we could take $|z|\to z\,{\rm
tanh}(10z)$. For this particular smoothing in $d=6$ with $k=1$,
Figure~\ref{fig:6d} shows the potential $V(z_1,z_2)$ appearing in the
equation for the gravitational fluctuations \eqref{mdschreq}.
\begin{figure}
\begin{center}
\leavevmode \epsfxsize=7cm \epsfbox{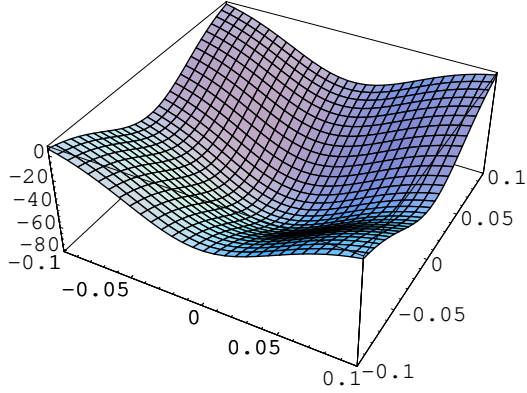}
\end{center}
\caption{\small Close-up of the potential near the brane intersection.}
\label{fig:6d'}
\end{figure}

\subsection{A solvable example}

In general, the Schr\"odinger equation \eqref{mdschreq} in the case of intersecting
branes is fully $n$-dimensional and consequently rather complicated; 
however, in the case when
\EQ{
A(z)=\sum_{i=1}^n A^{(i)}(z^i)\ ,
}
it is separable by writing
$\psi(z)=\psi^{(1)}(z^1)\times\cdots\times\psi^{(n)}(z^n)$. 
In this case the solution represents $n$
intersecting $(d-2)$-branes where the intersection is four-dimensional.

For each $i$, $\psi^{(i)}(z^i)$ satisfies a one-dimensional
Schr\"odinger equation identical to \eqref{scheq} with $A(z)\to
A^{(i)}(z^i)$. Hence, we can immediately draw on
the results of Sec.~\ref{sec:domain} to establish the conditions under
which gravity is localized on the intersection. 
Basically we require that 
$\exp\big[-\tfrac32A^{(i)}(z^i)\big]$ falls off faster than $1/z^i$ as
$|z^i|\to\infty$, for each $i$, so that the zero-energy state
\EQ{
\psi_0(z)=\psi^{(1)}_0(z^1)\times\cdots\times\psi^{(n)}_0(z^n)\ ,
}
is localized in all the transverse directions and therefore
represents the four-dimensional graviton. In general, the spectrum consists of 
$2^n$ different sectors, depending on whether each $\psi^{(i)}(z^i)$ is the
normalizable wavefunction $\psi^{(i)}_0(z^i)$ or a continuum wavefunction
$\psi_m^{(i)}(z^i)$. So for example, when there are 2 transverse
directions there will be 4 sectors in the spectrum spanned by
\begin{alignat*}{2}
\text{(1)}\quad&\psi_0^{(1)}(z^1)\psi_0^{(2)}(z^2)\ ,&\qquad
\text{(2)}\quad&\psi_0^{(1)}(z^1)\psi_{m_2}^{(2)}(z^2)\ ,\\
\text{(3)}\quad&\psi_{m_1}^{(1)}(z^1)\psi_0^{(2)}(z^2)\ ,&\quad
\text{(4)}\quad&\psi_{m_1}^{(1)}(z^1)\psi_{m_2}^{(2)}(z^2)\ .
\end{alignat*}
The first state (1) corresponds to the four-dimensional graviton. The 
set of continuum states (4) contributes to Newton's Law as in
\eqref{cnl},
\EQ{
\delta_4U(r)\thicksim
M^{2-d}_*\int_{0}^\infty dm_1dm_2\, 
\frac{M_1M_2e^{-\sqrt{m_1^2+m_2^2}r}}{r}\psi_{m_1}^{(1)}(0)^2\psi_{m_2}^{(2)}(0)^2\ ,
}
The set of continuum states (2) or (3) are localized in one
direction and contribute as an integral over the single eigenvalue $m_2$ and
$m_1$, respectively, 
\EQ{
\delta_{2+3}U(r)\thicksim
M^{2-d}_*k\int_{0}^\infty dm_1\,
\frac{M_1M_2e^{-m_1r}}{r}\psi_{m_1}^{(1)}(0)^2+M^{2-d}_*k\int_{0}^\infty dm_2\,
\frac{M_1M_2e^{-m_2r}}{r}\psi_{m_2}^{(2)}(0)^2\ .
}

\subsection{Intersecting branes from scalar fields}\label{sec:intscalar}

In this section, we consider whether the conformally-flat intersecting branes
background can be generated from a single real scalar field.   
We argue that a single scalar
field cannot produce a general intersecting brane background, 
but this leaves open the possibility
that such backgrounds can be generated from more than one scalar field;
for instance a single complex scalar as in \cite{SeanMark}. In
addition, we will also see, as promised in \ref{sec:scalars}, that a
single real scalar field cannot produce a conformally-flat three-brane embedded in $d>5$.

We assume the metric has the form \eqref{confflat}, and there is a single
scalar field which generates the energy-momentum tensor, 
\begin{equation}
T_{\mu\nu}=\tfrac12\partial_\mu\phi(z)\partial_\nu\phi(z) - \tfrac12g_{\mu\nu}
\big(\tfrac{1}{2}\partial^\rho\phi(z)\,\partial_\rho
\phi(z)-{\cal V}[\phi(z)]\big)\ .
\end{equation}
The Einstein tensor $G_{\mu\nu}$ can be read off of \eqref{conformal},
and satisfies, for components transverse to the intersection,
\begin{equation}
G_{ii}=-G_{00}+\tfrac{d-2}2\big[\tfrac12(\partial_iA(z))^2+\partial_i^2A(z)\big]\
. 
\end{equation}

The diagonal components of the stress tensor satisfy a similarly simple 
relation, 
\begin{equation}
T_{ii}=-T_{00}+\big(\partial_i\,\phi(z)\big)^2\ , 
\end{equation}
so the particular combination $G_{ii}+G_{00}=\kappa^2(T_{ii}+T_{00})$ of components
of the Einstein equation relates derivatives of the scalar field to 
derivatives of the metric, independent of the form of the scalar
potential ${\cal V}(\phi)$:  
\begin{equation}
\tfrac12\kappa^2\big(\partial_i\,\phi(z)\big)^2=\tfrac{d-2}2
\big[\tfrac12\partial_iA(z)\partial_iA(z)+\partial_i^2A(z)\big]\ . 
\label{Eii+E00}
\end{equation}
The off-diagonal components of the Einstein equation are,
\begin{equation}
G_{ij}-\kappa^2T_{ij}=\tfrac{d-2}2
\big[\tfrac12\partial_iA(z)\partial_jA(z)+\partial_i\partial_jA(z)\big]-
\tfrac12\kappa^2\partial_i\phi(z)\partial_j\phi(z)=0\ ,
\end{equation}
or, using \eqref{Eii+E00}, 
\begin{equation}
\big[\partial_i\partial_je^{A(z)/2}\big]^2=\big[\partial_i^2e^{A(z)/2}\big]\big[
\partial_j^2e^{A(z)/2}\big]\ . 
\label{E45} 
\end{equation}
This constrains the type of metric which can be obtained by a
gravitating scalar field.  In particular, the solution of \eqref{E45} is,
\begin{equation}
A(z)={\cal A}\big(\sum_{i=1}^na_i\,z^i\big)\ , 
\end{equation}
for some constant coefficients $a_i$.  By a redefinition of coordinates, this 
can always be recast in the form $A(z)={\cal A}(z^1)$, which preserves a
$(d-1)$-dimensional Lorentz symmetry and hence can create only
a single $(d-2)$-brane in $d$ dimensions. 
Therefore, it seems to be the case that additional
fields are required to create intersecting branes in the presence of
gravity. In addition, we conclude that a single real scalar field cannot produce a
$p$-brane in $d>p+2$.

\section{Conclusions}\label{sec:con}

We have studied general features of gravity in domain wall backgrounds.  
For general domain wall type metrics we have found the conditions for
there to be localized gravity on the domain wall or on the intersection
of domain walls.  It turns out that it is possible in generalizations of the
Randall-Sundrum scenario for the graviton zero mode to be non-normalizable.
It is precisely in that case that the effective gravitational coupling
on the domain wall vanishes, and that the Kaluza-Klein modes become relevant at
long distances.  We have seen several illuminating examples in which 
smooth
domain wall backgrounds are created by scalar fields, and in which the 
domain walls are created by non-dynamical sources in the absence of
fields other than the graviton.  
Negative tension branes, such as appear in the RS solution to the
hierarchy problem,  can be studied in this formalism.
We have studied intersecting domain walls in the absence of fields other than
gravity, and we argued that a single scalar field is not sufficient to 
produce intersecting domain walls.  We commented on resonant modes in the
continuum of Kaluza-Klein modes, and argued that they are unimportant in 
smooth versions of the RS scenario.

One issue that is worth commenting on is the fate of the
Lykken--Randall scenario for the solution of the hierarchy problem~\cite{LR}, 
in the context of thick branes. In this
scenario we consider again the general five-dimensional backgrounds
\eqref{metfive} but now associate the matter fields that describe our
world with another brane---the ``TeV brane''---located at a point
$z_0$ in the transverse space. In order to create the necessary
hierachy of scales we need $z_0$ such that
\EQ{
M_{\rm Pl}^2e^{-2A(z_0)}\sim{\rm TeV}^2\ ,
}
where $M_{\rm Pl}$ is the Planck mass. In the context of the
backgrounds discussed in Sec.~\ref{sec:domain}, this requires that
$z_0$ is much larger than $k$ (since $k\sim M_{\rm Pl}$). In other
words, the TeV brane sits at a place where we may approximate the
potential $V(z)$ by its asymptotic form \eqref{potas}.
Since the relevant dynamics now takes place at
$z=z_0$ we have to re-assess the effects of the KK modes. In
particular, the effective Newton's constant on the brane is now given by
\EQ{
G_N \sim
\frac{M^{2-d}_*\hat\psi_0(z_0)^2}{\langle\hat\psi_0|\hat\psi_0\rangle}\ .
}
This modifies the effects of the KK modes; in the case
when the continuum starts at $m=0$, the corrections from the KK modes 
\eqref{cnl} are modified to \cite{LR}
\EQ{
U(r)\thicksim
G_N\frac{M_1M_2}r\Big[1+\int_{0}^\infty dm\, 
\frac{M_1M_2e^{-mr}}{r}\Big(\frac{\psi_m(z_0)}{\hat\psi_0(z_0)}\Big)^2\Big]\ .
\label{mcnl}
}
In the above, the zero-energy wavefunction $\hat\psi_0(z)$ must be
normalized to unity $\langle\hat\psi_0|\hat\psi_0\rangle=1$.
As $z_0$ increases then one might worry that the ratio $\psi_m(z_0)/\hat\psi_0(z_0)$
becomes larger and the effect of the KK modes is less suppressed and
there would be large corrections to Newton's Law. Actually, as
explained in \cite{LR}, this is not the case. Since $z_0$ lies in the
tail of the potential \eqref{potas}, we can use the exact form of the
solution in terms of Bessel functions \eqref{KKwf} to assess the
magnitude of the corrections. For small enough $m$ the first term in
\eqref{KKwf} dominates, and one finds using the expansion
\eqref{expan} (taking into account the correct normalizations on the wavefunctions)
\EQ{
\frac{\psi_m(z_0)}{\hat\psi_0(z_0)}\simeq\frac{\psi_m(0)}{\hat\psi_0(0)}
\propto m^{\alpha-1}\ ,
\label{frett}
}
and so the suppression is unchanged from the situation where matter
fields are located at $z=0$. Small enough $m$ in this context means
that we can approximate $\psi_m(z_0)$ by the first term in
\eqref{expan}.  

The discussion above, where the matter fields are located at $z=z_0$
but are still point-like in the fifth dimension, also suggests that in
a more realistic scenario, where the matter fields are smeared out in
the fifth dimension, the corrections to Newton's law will also be
similarly suppressed and our previous insistence that the matter
sources were point-like and located at $z=0$ was not overly simplistic.

\acknowledgments

We thank Martin Gremm for sharing an early version of \cite{Gremm} with us 
prior to publication. We also thank Tanmoy Bhattacharya,
Fred Cooper, Dan Freedman, 
Michael Graesser, Martin Gremm, Salman Habib, Juan
Maldacena, Asad Naqvi, Michael Nieto and
Erich Poppitz for useful discussions, and Shanta de Alwis and
Christophe Grojean for comments and corrections. 
C.C. is an Oppenheimer Fellow at the 
Los Alamos National Laboratory.
C.C., J.E. and T.J.H. are supported 
by the US Department of energy under contract W-7405-ENG-36.
Work of Y.S. is supported by NSF grant PHY-9802484

\end{document}